\documentclass[]{article}
\usepackage{arxiv}
\usepackage{setspace} 
\usepackage[utf8]{inputenc} 
\usepackage[T1]{fontenc}    
\usepackage{hyperref}       
\usepackage{url}            
\usepackage{booktabs}       
\usepackage{amsfonts}       
\usepackage{nicefrac}       
\usepackage{microtype}      
\usepackage{lipsum}		
\usepackage{graphicx}
\usepackage{natbib}
\usepackage{doi}

\usepackage{amsmath}
\usepackage{enumitem}
\usepackage{graphicx}
\usepackage{tabularx}
\usepackage{mathtools}
\usepackage{amsmath}
\usepackage{amsfonts}

\usepackage{multirow}
\usepackage{soul}
\usepackage{sidecap}
\usepackage{algorithm}
\usepackage{algpseudocode}
\usepackage{adjustbox}
\usepackage{float}

\usepackage{titlesec}
\titlespacing{\section}{0pt}{\parskip}{-\parskip}
\titlespacing{\subsection}{0pt}{\parskip}{-\parskip}
\titlespacing{\subsubsection}{0pt}{\parskip}{-\parskip}

\usepackage{amsthm}
\theoremstyle{definition}
\newtheorem{definition}{Definition}[section]

\title{A REINFORCEMENT LEARNING APPROACH TO
RESOURCE ALLOCATION IN GENOMIC SELECTION}

\author{
  Saba Moeinizade\thanks{Corresponding author, email: sabamz@iastate.edu}\ , \ Guiping Hu,  \ and  \ Lizhi Wang \\
  Industrial and Manufacturing Systems Engineering Department\\
  Iowa State University\\
}

\renewcommand{\shorttitle}{\textit{arXiv}}
\hypersetup{
pdftitle={A template for the arxiv style},
pdfsubject={q-bio.NC, q-bio.QM},
pdfauthor={David S.~Hippocampus, Elias D.~Striatum},
pdfkeywords={First keyword, Second keyword, More},
}
\begin{document}
\maketitle
\section{Abstract}
Genomic selection (GS) is a technique that plant breeders use to select individuals to mate and produce new generations of species. Allocation of resources is a key factor in GS. At each selection cycle, breeders are facing the choice of budget allocation to make crosses and produce the next generation of breeding parents. Inspired by recent advances in reinforcement learning for AI problems, we develop a reinforcement learning-based algorithm to automatically learn to allocate limited resources across different generations of breeding. We mathematically formulate the problem in the framework of Markov Decision Process (MDP) by defining state and action spaces. To avoid the explosion of the state space, an integer linear program is proposed that quantifies the trade-off between resources and time. Finally, we propose a value function approximation method to estimate the action-value function and then develop a greedy policy improvement technique to find the optimal resources.
We demonstrate the effectiveness of the proposed method in enhancing genetic gain using a case study with realistic data.
\section{Introduction}
Over the past decades breeding methods have evolved from traditional phenotype-based selection to marker-assisted selection methods. Genomic selection (GS), which was initially proposed by \cite{Meuwissen2001}, is a special form of marker assisted selection that estimates the effects of genome-wide markers in a training population consisting of genotyped and phenotyped individuals. Different statistical and machine learning models are proposed to develop prediction models based on the genotypic and phenotypic data of the training population \citep{1pryce2011genomic,neves2012comparison,2chen2014impact,3li2015novel,4dong2016comparative,7liu2017application,8crossa2017genomic,9montesinos2018multi,6liu2018genomic}. Then, the prediction model is used to derive the genomic estimated breeding values (GEBVs) for all individuals of the breeding population (BP) from their genomic profile by calculating the sum of the estimated marker effects. Given the genotype information and the estimated marker effects of individuals in a breeding population, there are different decisions that should be made within each breeding cycle. These decisions include selection, mating, and resource allocation which must be made in every generation with the objective of continuously improving individuals subject to deadline constraints \citep{moeinizade2018stochastic,moeinizade2020complementarity, MoeinizadeMTLAS,moeinizade2021look}. 

Recently, \cite{moeinizade2019optimizing} presented the look-ahead selection (LAS) method to optimize selection and mating strategies with a time-dependent approach. A new technique was invented to anticipate the consequences of selection and mating decisions through several generations, which was achieved by quantitatively taking into account recombination frequencies. Recombination, the main source of uncertainty in reproductive biology, is the phenomenon that occurs during meiosis and creates different combinations of alleles in the resulting gametes \citep{10lobo2008thomas}. 
In \cite{moeinizade2019optimizing}, we conducted a case study using realistic maize data and compared LAS with other published selection methods. Simulation results suggested the superiority of LAS to other selection methods.
However, the LAS method was unable to optimize resource allocation decisions, e.g., how should the budget be distributed over time? Should it be spent evenly or should more investment be made in earlier generations before genetic diversity deteriorates? how many crosses should be made and how many progeny should be produced? These resource allocation decisions should be optimized systematically, given the cost of making a cross and genotyping progeny, under budget and deadline constraints, considering the uncertainty in recombination in each generation.

In this study, we develop a reinforcement learning-based algorithm to automatically learn to allocate resources across different generations of breeding. The proposed new method integrates the LAS approach in a reinforcement learning framework. The LAS method is capable of anticipating the consequences of the selection and mating decisions under uncertain recombination events efficiently and accurately, whereas the reinforcement learning framework is capable of making a trade-off between cost and time which is necessary to make resource allocation decisions.

Reinforcement learning (RL) is one of the most important research directions of machine learning, which has been widely used in different fields like social sciences, natural sciences, and engineering and has significantly impacted the development of Artificial Intelligence (AI) over the last years \citep{11dayan2008reinforcement}. \cite{12sutton2018reinforcement} define Reinforcement learning as learning what to do —how to map situations to actions— so as to maximize a numerical reward signal. The main characters of RL are the agent and the environment. The environment represents the outside world to the agent and the agent interacts with the environment by taking actions and receiving a reward signal. The goal of the agent is to maximize the cumulative reward, named return. To do that, the agent should learn the optimal policy which is an optimal strategy to behave in the environment.

RL problems can be formulated mathematically in the framework of Markovian Decision Processes (MDPs) by defining states, actions, transition probabilities, and rewards \citep{18szepesvari2010algorithms}. The transition and reward functions in MDPs are called the model of environment. A known MDP can be solved by dynamic programming which relies on simplifying a complicated problem by breaking it down into simpler sub-problems in a recursive manner \citep{bellman1966dynamic}. However, we often do not have the transition and the rewards of the MDP. This class of problems with unknown MDPs are called model-free. While model-based methods rely on planning as their primary component, model-free methods rely on learning \citep{12sutton2018reinforcement}. Model-free methods can be applied to both prediction and control problems. In model-free prediction, the goal is to estimate the value function of an unknown MDP where as model-free control aims at optimizing the value function. The value function represents how good it is for an agent to be in a given state.

In recent years, different solution methods have been proposed to solve model-free RL problems \citep{13mnih2013playing,19schulman2015trust,20schulman2015high,14hausknecht2015deep,16van2015deep,15wang2016dueling,21schulman2017proximal,22heess2017emergence,23tucker2018mirage}. 
These solution methods include two main types of algorithms, value-based and policy-based. Value-based algorithms iteratively update the value of a state to finally learn an optimal policy. Policy-based algorithms learn a parameterized policy that can select actions without consulting a value function.

Q-learning, a value-based RL algorithm, is one of the most popular solution methods in reinforcement learning. This algorithm uses Q-values (an estimation of how good it is to take an action at a given state) to iteratively improve the behaviour of the learning agent \citep{watkins1992q}. However, for large-scale problems with an enormous number of state-action pairs, it is difficult to explicitly store all the Q-values. To overcome this challenge, function approximation methods are used where value function is represented by mapping a state description to a value \citep{gosavi2009reinforcement,kaelbling1996reinforcement, arulkumaran2017deep}. Many implementations of RL in real-world problems have used neural networks as function approximators \citep{13mnih2013playing,14hausknecht2015deep,16van2015deep,15wang2016dueling}. One of the examples is the achievement of AlphaGo in 2016, where a deep Q-network was implemented and trained to predict total reward \citep{17silver2016mastering}. 
Other approximation methods including kernel methods, nearest-neighbor algorithms, and decision trees can be used to estimate the Q-values \citep{friedman2001elements, chapman1991input,howe1998decision}. Policy gradient algorithms learn in a more robust way by approximating policy and updating it according to the gradient of expected reward with respect to the policy parameters \citep{sutton1999policy} without the need to construct a value function.

In this study, we propose a value-based algorithm with function approximation and introduce a backward greedy policy approach with respect to the estimated values (i.e., the policy that selects the action with highest estimated value in each state). The idea of the backward approach is to learn the optimal action in a backward manner starting from the final generation to the first generation given that the optimal strategy in the final generation is allocating all remaining resources. 
In the remainder of this paper, we formulate the resource allocation problem in an RL framework, discuss the solution methods and finally present a case study to compare our proposed allocation strategy with even allocation using computer simulation. 
\section{Methods}
In this section, we first define the genomic selection resource allocation problem and then formulate the proposed problem mathematically in the context of Markov Decision Process (MDP), where reinforcement learning algorithms can be used. Finally, we provide a solution method to solve the proposed MDP and find the optimal policy. 

\subsection{Problem Definition}
A classical plant breeding process starts with an initial population and iteratively goes through the selection and reproduction steps until getting the final population. In addition to the selection decisions, in each generation, the breeder should decide how to allocate resources (i.e., the number of crosses to be made and the number of progeny to be produced from each cross). The focus of this study is optimizing the resource allocation strategy in a breeding program. 

Let $G_{t} \in \mathbb{B}^{L \times M \times N}$ represent the genotype of the population at generation $t$, where $L$ is the total number of alleles, $M$ indicates the ploidy of the plant ($M=2$ for diploid species) and $N$ is the total number of individuals in the population. For all $l$, let $\beta_l$ denote the additive effect of allele $l$, which is assumed to have been reasonably estimated.
 Given $\beta$ and $G$, the look-ahead selection (LAS) algorithm can optimize the selection and mating steps with a time-dependent approach \citep{moeinizade2019optimizing} by maximizing the expected GEBV of the best offspring in the terminal generation ($T$) where GEBV of an individual can be calculated as the sum of all marker effects across the entire genome.
 
 Let the cost of producing one progeny be one unit of budget. Then, spending $b$ units of resources in the current generation will produce $b$ progeny.
 Given a fixed amount of total budget of $B_{0}$ units of resources over $T$ generations, the goal is to find the optimal budget or population size for each generation, ($b_1,b_2,...,b_T$), in order to maximize the performance of individuals in the final generation. Similar with selection and mating, resource allocation decisions should be made in a dynamic manner after observing the genotype of progeny from previous generations, while considering the total budget constraint over $T$ generations: $\sum_{t=1}^{T} b_{t}\leq B_{0}$.

\subsection{Problem Formulation} \label{PF}
Here, we present the MDP formulation for the genomic selection resource allocation problem.
An MDP process is described by a finite set of states (S), a finite set of actions (A), transition probabilities (T), and a reward function (R). 
Due to the stochastic nature of the environment, in this problem, we cannot derive the transition probabilities and the reward is delayed until the terminal generation. Hence, we use learning to understand the behavior of the environment by simulating different scenarios of resource allocation (section \ref{PS}). In this section, we define the state and action spaces for the MDP.

\subsubsection{State Space} 
 To capture the full information in each generation, the population genotype would be necessary to define the state space, but it would make the resulting model unsolvable. For example, for a small population of $200$ individuals and only $10,000$ pairs of genes, the dimension of the state space would be $~3^{2,000,000}$ with each pair of genes taking three possible combinations of two variants of alleles (AA, aa, or Aa). To avoid formidable dimensions, we need to simplify the state space by presenting a compact definition that captures the important information by considering the current genetic value of the population and quantifying the trade-off between time and resources.

 At generation $t$, we define the state by $\left(g^{\max }_t, C_t, B_{t-1}\right)$, where $g^{\max }_t$ is the highest GEBV of the $N$ individuals at generation $t$ calculated as follows:

\begin{equation}
g^{\max}_t=\text{max}_{n \in  \{1,2,...,N\}}\left(\sum_{l=1}^{L} \sum_{m=1}^{2} G^{l, m, n}_t \beta^{l}\right)
\end{equation}

 In this state space definition, $C_t \in \mathbb{R}^{K \times M}$ measures the specific combining ability, and $B
 _{t-1}$ is the available budget to be spent in generations $t$ to the end. Specifically, $C^{k, m}_t$ is the highest possible GEBV of a gamete that could be assembled from $G_t$ using at most $m$ individuals with recombination events that are more likely than $p^{k}$, where $p \in (0,1)$ is an adjustable parameter, depending on the sensitivity of recombination frequency and available resources.  
The value $C^{k, m}_t$ measures the potential of the genotype $G_t$ to create a gamete with the highest possible GEBV subject to resource and time constraints. The first dimension $k$ reflects the constraint of probabilistic recombinations afforded by remaining resources, and the second dimension $m$ indicates the number of founding parents that the gamete needs to collect alleles from, which would require $\left\lfloor\log _{2} m\right\rfloor$ generations of breeding. Value $C_t^{k, m}$ can be obtained using the following integer linear program. 

\begin{eqnarray}
\max _{x, y, z} & C^{k, m}_t=\sum_{i} \sum_{c} \sum_{j} \beta_{i} G^{i, c, j}_t x_{i, c, j}  \label{const0}\\ 
\mathrm{s.t.} & \sum_{j}\left(x_{i, 1, j}+x_{i, 2, j}\right)=1 & \forall i \label{const1}\\
&\sum_{i}\left(x_{i, 1, j}+x_{i, 2, j}\right) \leq L y_{j} & \forall j  \label{const2}\\
&x_{i, c, j}-x_{i+1, c, j} \leq z_{i} & \forall i, c, j \label{const3} \\
&\sum_{j} y_{j} \leq m \label{const4}\\
&\prod_{i}\left(\frac{r_{i}}{1-r_{i}}\right)^{z_{i}} \geq p^{k} \label{const5}\\
& x, y, z \text { binary } \label{const7}
\end{eqnarray}

Here, $x_{i, c, j}$ is a binary variable that indicates whether allele $(i, c, j)$ is selected $\left(x_{i, c, j}=1\right)$ or not $\left(x_{i, c, j}=\right.$ 0 ) to assemble the gamete, $y_{j}$ is a binary variable that indicates whether individual $j$ is used $\left(y_{j}=1\right)$ or not $\left(y_{j}=0\right)$, and ${z}_{i}$ is a binary variable that indicates whether there is a recombination between loci $i$ and $i+1$ $\left(z_{i}=1\right)$ or not $\left(z_{i}=0\right)$.

The objective value \eqref{const0} is the maximum possible GEBV of a gamete that can be assembled from the current population. Constraint \eqref{const1} ensures selection of one chromosome for each locus in an individual to assemble the gamete.
 Constraint \eqref{const2} requires that no alleles from unselected individuals can be used to assemble the gamete.      
 Constraint \eqref{const3} detects whether a recombination is necessary between loci $i$ and $i+1$.   
 Constraint \eqref{const4} limits the selection of at most $m$ parents among all individuals. Finally, constraint \eqref{const5}
 requires that the likelihood of necessary recombinations be no less than $p^k$, which equivalently limits the amount of resources needed to afford such recombination events.  
 Take for example, for very small $r_{i}$ values, the value of $\left(\frac{r_{i}}{1-r_{i}}\right)$ becomes very small, hence we need more resources (larger $k$) to make that recombination happen.

\subsubsection{Action Space} 
The decision maker should take an action in each generation and decide the amount of resources and the selection strategy for that generation. In \cite{moeinizade2019optimizing}, we demonstrated the effectiveness of look-ahead selection (LAS) against conventional selection methods. Here, we focus on optimizing the resource allocation and use LAS algorithm to determine the selection strategy.

A common way of obtaining approximate solutions for continuous action spaces is to discretize the action space. In a discrete action space, the agent decides which distinct actions to perform from a finite action set \citep{masson2016reinforcement}.
In this study, we discretize the action space. Specifically, we assume allocating $b$ amount of resources in one generation is equal to producing $b$ total number of progeny in that generation (making one progeny costs one unit of budget). Hence, action is a discrete value representing the number of progeny in the population.

We define the action space as time dependent set of $(b_1,b_2,...,b_T)$ values such that $\sum_{t=1}^{T} b_{t}= B_{0}$ where $b
_t$ is the amount of resources to spend in generation $t$, $T$ is the total number of generations, and $B
_{0}$ is the amount of total budget.

\subsection{Proposed Solution Technique} \label{PS}
\begin{figure}[H] \centering
\includegraphics[width=14 cm]{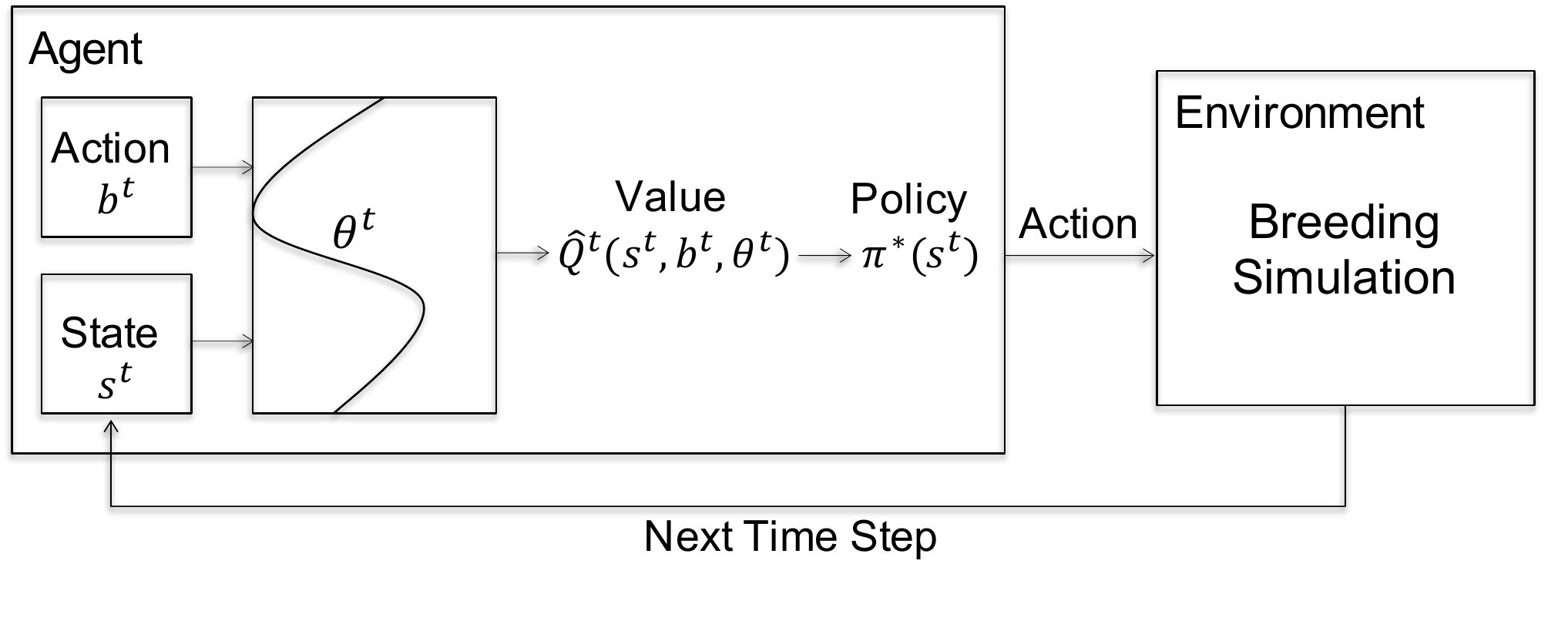}
\caption{The reinforcement learning system representation. The environment is the breeding simulation which provides the next state to the agent. At each time step, the value of the current state is evaluated using a pretrained nonlinear function for a given action. Finally, the policy function determines the optimal action which is passed to the simulation.}
\label{framework}
\end{figure}

Figure \ref{framework} represents the reinforcement learning system. As shown in this figure, there are two main components in the system: agent and environment. The goal of the agent is to find the optimal policy (i.e., the optimal action to take at time $t$) where policy is defined as a function mapping states to the actions ($\pi: S \xrightarrow{} A$). 
The environment is the breeding simulation which provides  the next state to the agent. The agent evaluates the value of performing action $b$ in state $s$ using a pretrained value function and then acts greedily to find the best policy. 
The agent then decides to take the optimal action at time $t+1$ and this process continuous until reaching the deadline. 

Suppose we have an MDP defined as state-action pairs and given some policy $\pi$. First, we predict the value function by constructing the action-value function $\hat{Q}(s,b,\theta)$ to represent the objective value for a given state-action pair. Then, we can predict the value of a state given all possible actions and find the optimal policy, $\pi^{*}(s)$, that maximizes the action-value function. Section \ref{VFA} describes the value function approximation technique and section \ref{GPI} elaborates on policy improvement.

\subsubsection{Value Function Approximation} \label{VFA}
The value function demonstrates how good each state and/or action is by calculating the expected cumulative reward in long-term. In this problem, the immediate rewards are considered to be zero and the objective is to maximize the genetic gain in final generation. Hence, the value here represents the GEBV of the best offspring in the final generation, where GEBV of individual $n$ is calculated as the sum of effects across the entire genome (GEBV$(n)=\sum_{l=1}^{L} \sum_{m=1}^{2} G^{l, m, n} \beta^{l} $).

The simplest way of representing a value function is by the use of a lookup table, with the values of each state-action pair stored. However, when the state-action spaces are large, storing and retrieving values become a problem, as it takes up large amounts of computational resources. To solve this problem, function approximators can also be used instead of a lookup table for representing value functions, thereby limiting the memory being used and speeding up the learning process. Therefore, to estimate the value function, $\hat{Q}(s,b,\theta)$, efficiently, we should use a function approximation method (e.g., nonlinear regression, support vector machine, decision tree based models and neural network). 

The parameters ($\theta$) need to be learned for each time period, $t$, separately. 
 We employ a backward approach by optimizing resources from the final generation to the first generation. Given that the objective is to maximize the maximum GEBV in the target generation, ($g^{\max}_T$), the optimal strategy in the final generation is to allocate all the remaining budget (${b}_T^*=B_{T-1}$, where $B_{T-1}$ is the remaining resources for the final generation).

 To find optimal budget, ${b}_t^*$, for earlier generations $t \in \{1, ..., T-1\}$, we take advantage of simulation to learn how different budget allocation scenarios impact the final performance by generating learning data as described in algorithm \ref{AT}. This algorithm presents data collection process for a given generation, $\tau$, which goes backwards from $T-1$ to 1. For generation $\tau$, we record state-action pairs, $(g^{\max}_\tau,C_\tau,B_\tau)$ and $b_\tau$, and the objective value, $g^{\max}_T$. Then, we estimate the value function to map state-action pairs to the objective value. We first define the three functions used in algorithm \ref{AT} for data generation and then discuss the value function approximation technique.

\begin{definition}
The selection function is defined as follows: $[S]=\texttt{Select}(G_{t-1},r,n,b_t)$. The input parameters are the population genotype at generation $t-1$, $G_{t-1} \in \mathbb{B}^{L \times 2 \times N}$, recombination frequency vector, $r \in [0,0.5]^{N-1}$, the number of crosses $n$, and amount of resources for generation $t$, $b_t$. Note that the resources correspond to the progeny population size. The output $ S=\begin{bmatrix}
s_{1,1} & s_{1,2} & b_t^{1}\\
s_{2,1} & s_{2,2} & b_t^{2}\\
    &  \dots  &       \\
s_{n,1} & s_{n,2} & b_t^{n}\\
\end{bmatrix}$ contains the indices of selected parents in the breeding population and the number of progeny produced from each cross (here, row $\begin{bmatrix}s_{i,1} &  s_{i,2} & b_t^i\end{bmatrix}$ means that the $i^\text{th}$ cross in generation $t$ is made using parents $s_{i,1}$ and $s_{i,2}$, which produces $b_t^i$ progeny. The numbers of progeny satisfy the budget constraint that $\sum_{i=1}^{n} b_t^{i}=b_t$).
\end{definition}

\begin{definition}
The reproduction function is defined as follows: $[G_{t}]=\texttt{Reproduce}(G_{t-1},S,r)$. The input parameters are the  population genotype at generation $t-1$, $G_{t-1} \in \mathbb{B}^{L \times 2 \times N}$, selection matrix, $S$, and the recombination frequency vector, $r \in [0,0.5]^{N-1}$. The output is the genotype of the progeny population. The genetic information are inherited from parents to progeny according to the inheritance distribution defined in \cite{Han2017}.
\end{definition}

\begin{definition}
The action function is defined as follows: $[b_t]=\texttt{Action}(t,T,B_{t-1},A)$. The input parameters are the current generation, $t$, total number of generations, $T$, the available resources to be spent in generations $t$ to the end, $B
_{t-1}$, and possible set of actions, $A$. The output is resources or progeny size for generation $t$.
We choose an action randomly from a finite set of values $\Tilde{a}\in A, \ A=\{a_1,a_2,...,a_k\}$. We produce at least $\alpha$ progeny for each generation ($\alpha=\min_{i=1}^{k} a_i$).
Therefore the output, $b_t$, can be calculated as follows: $b_t=\text{min}(\Tilde{a}, B_{t-1}-\alpha \times (T-t))$.
\end{definition}

Generating learning/training data in complex stochastic environments can be time consuming. Neural networks usually need more training data and thus are not the best approach here since the learning process is considerably time-consuming. After exploring three function approximotors including generalized additive model (gam), support vector machine (SVM), and random forest (RF), we decided to choose random forest considering both efficiency and computational time. The inputs to the random forest model are the data generated using algorithm \ref{AT} with $km+3$ features including the maximum current GEBV ($g^{\max}_t$), the potential matrix ($C_t$), remaining budget ($B_{t-1}$), and action ($b_t$) where $k$ and $m$ are the dimensions of the potential matrix. The output is the GEBV of best individual in the final generation, $g^{\max}_T$.

\begin{algorithm}[!htb] \label{AT}
\begin{algorithmic} 
\caption{Learning data generation}
\State \text{Start with initial population $G_{0}$ and total budget $B_{0}$}
   \For{$t:=1$ to $\tau-1$}
        \State \texttt{$b_t=\text{Action}(t,T,B_{t-1},A)$}
        \State \texttt{$[S]=\text{Select}(G_{t-1},r,n,b_t)$}
        \State \texttt{$[G_{t}]=\text{Reproduce}(G_{t-1},S,r)$}
   \EndFor
    \For{$b_{\tau}\in A$}
    \For{$t:=\tau$ to $T$}
        \If{$t = \tau$}
        \State \text{Record $(g^{\max}_\tau,C_\tau,B_\tau)$ and $b_\tau$}
        \Else
        
        \If{$t = T$}
        \State \texttt{$b_t=B_{T-1}$}
        \Else
        \State \texttt{$b_t=\text{argmax}_{b\in A} \hat{Q}_{t}(s, b, \theta)$}
        \EndIf
        
        \EndIf
        \State \texttt{$[S]=\text{Select}(G_{t-1},r,n,b_t)$}
        \State \texttt{$[G_{t}]=\text{Reproduce}(G_{t-1},S,r)$}
        \State \text{Record $g^{\max}_T$}
    \EndFor
    \EndFor
    \end{algorithmic}
\end{algorithm}

\subsubsection{Greedy Policy Improvement} \label{GPI}
The ultimate goal of the agent is to find an optimal policy $\pi^{*}$ that maximizes the value function.
After learning the value function, we employ a greedy approach to improve the policy by selecting the action with the highest estimated value in each state. 
Let $\hat{Q}_{t}(s,b,\theta)$ represent the approximated value function for each generation except final. We can calculate the optimal policy for all generations from $1$ to $T-1$ as follows:

\begin{equation}\label{eqpolicy}
\pi^{*}_{t}(s)=\text{argmax}_{b\in A} \hat{Q}_{t}(s, b, \theta),\  \forall t \in \{1,2,..., T-1\}
\end{equation}

Moreover, the optimal policy in the final generation is to allocate all the remaining budget. Therefore $\pi^{*}_{T}(s)=B_{T-1}$, where $B_{T-1}$ presents the remaining resources for the final generation.
\section{Results}

\subsection{Simulation Settings}
 The genotypic data, marker effects and recombination rates are based on \cite{moeinizade2019optimizing}. The genotypic data contains genotypes of 369 maize inbred lines consisting of $L=1.4\text{M}$ SNPs distributed across ten maize chromosomes. To reduce the dimension, we define haplotype blocks. The resulted data has $L=10,000$ markers.

Let's assume there exist 5 total generations of breeding ($T=5$) and the amount of total budget is 1000. We consider seven possible action values as follows: $A=\{50,100,150,200,250,300,350\}$. 
Note that the amount of budget in each generation indicates the total number of progeny produced in that generation. Additionally, we consider that no more than 10 crosses are made in each generation.

\begin{figure}[!htb] \centering \label{Diagram}
\centering
\includegraphics[width=12 cm]{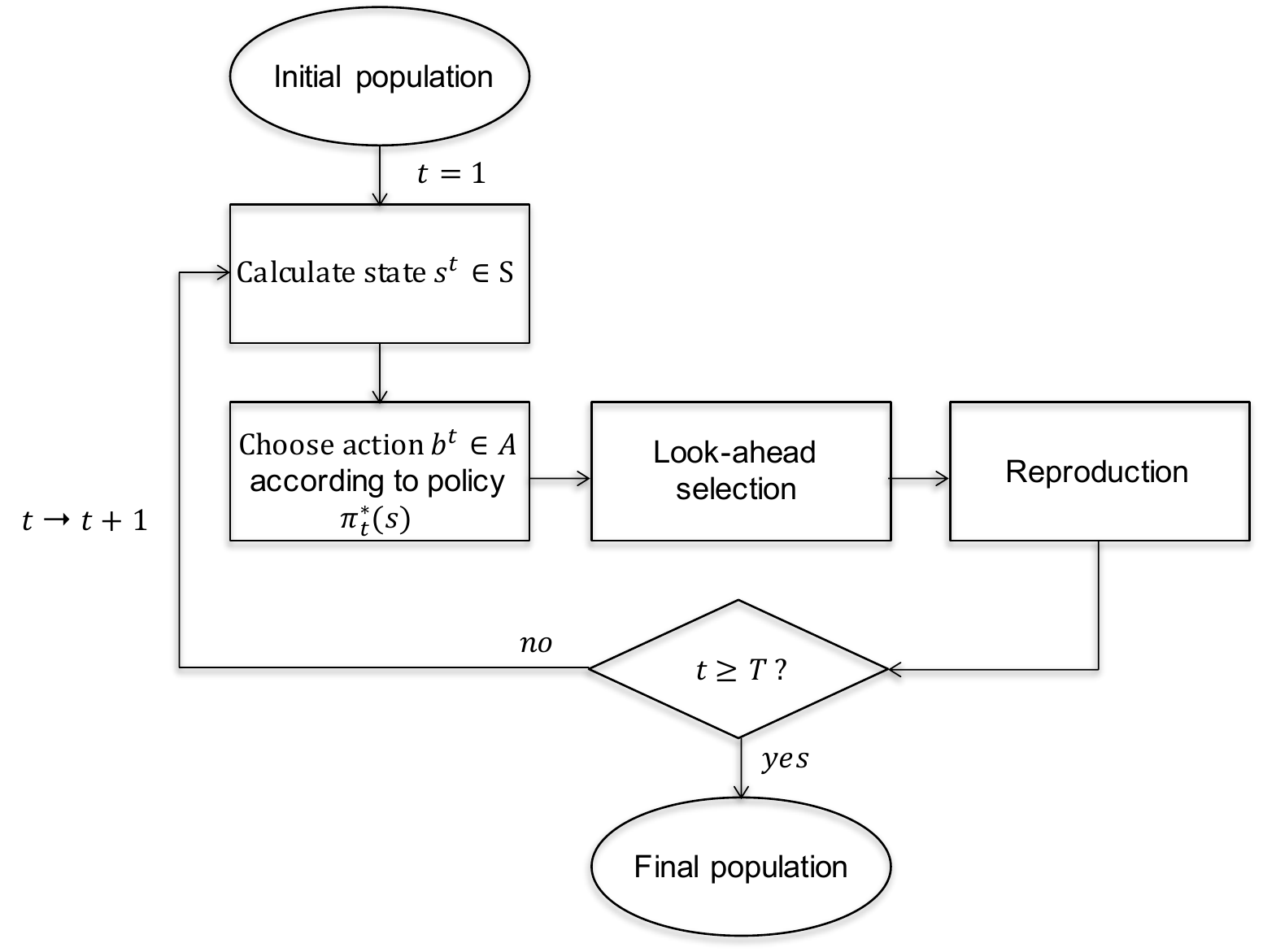}
\caption{The simulation flowchart. The process starts with the initial population and goes through resource allocation, selection and reproduction steps until getting to the deadline. To find the optimal resources, we first calculate the current state and then take the action with highest value according to the optimal policy. The action represents the number of progeny to be produced for that generation. }
\end{figure}

\begin{figure}[!htb] \centering
\includegraphics[width=3.5 in]{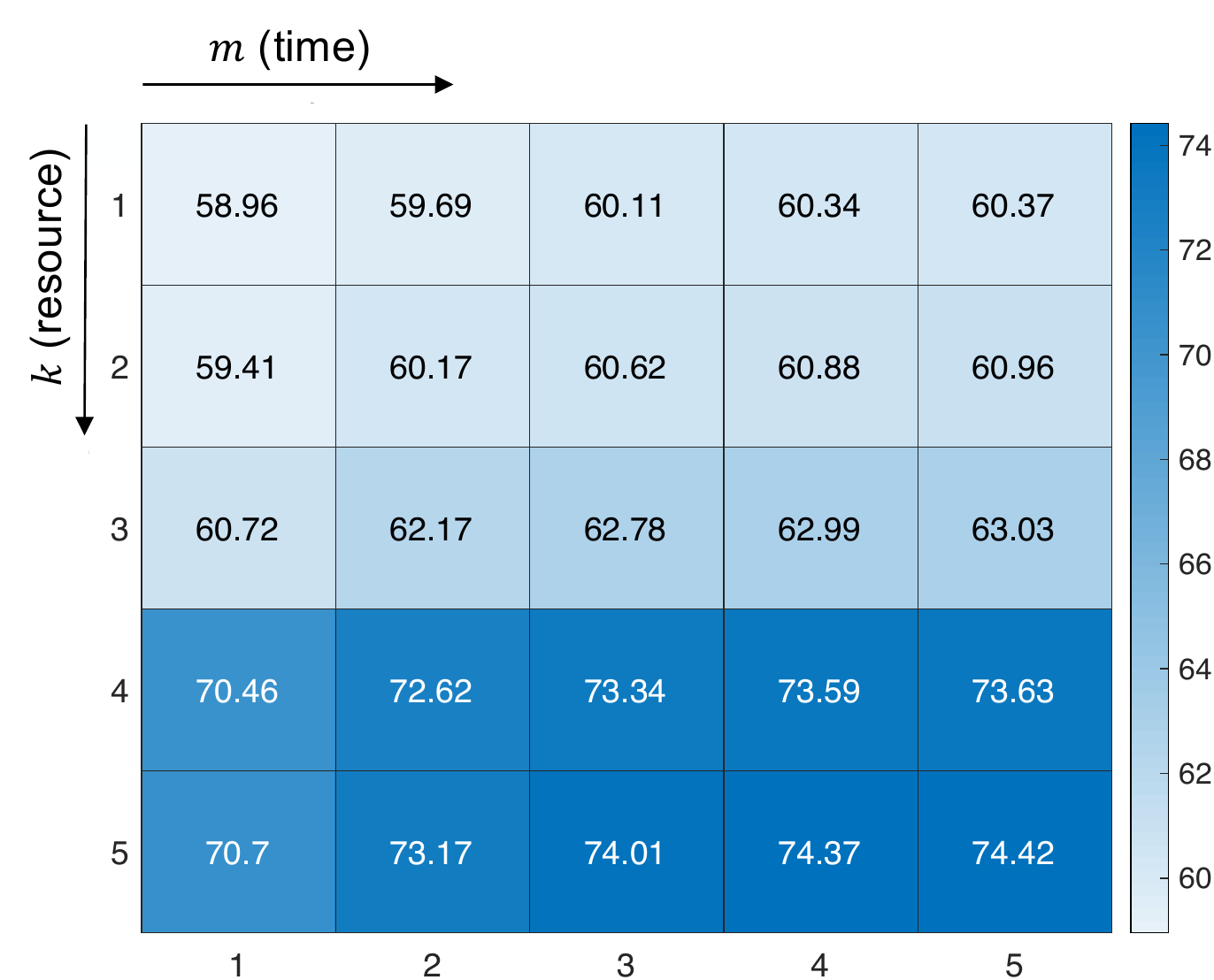}
\caption{Heat map for one sample $C$ matrix where $k=5$, and $m=5$. Each square demonstrates the best achieved GEBV value considering different levels of time and resources. The bottom right square has the highest potential GEBV value since it considers having the most time and resources.}
\label{heatmap}
\end{figure}

Figure \ref{Diagram} demonstrates the simulation flowchart. We start with the initial population by randomly choosing 200 individuals out of 369.  The state is observed by calculating the tuple $\left(g^{\max}_t, C_t,  B_{t-1}\right)$. The $C_{t}$ matrix is generated by solving the optimization problem presented in \eqref{const0}-\eqref{const7}. Here we choose $k=5$, and $m=5$. These are parameters and can be changed according to the data and time required for solving the optimization problem. Figure \ref{heatmap} shows the heat map for one sample $C_{t}$ in a simulation. The x-dimension of this plot is representing the possibility of combining more alleles from multiple individuals in case of having more time. Moreover, the y-dimension is representing the possibility of allowing more recombination to happen in case of having more resources. As expected, the performance becomes better towards the right and bottom of the plot by considering more time and resources. In addition, this C matrix indicates that given the current generation, potential genetic gain is more sensitive to resources than to time constraint, which is helpful for the reinforcement learning algorithm to make resource allocation decisions.

Next, the optimal policy will be calculated using the current generation action-value function in a greedy approach. 
Then, candidates are selected according to look-ahead selection as parents to produce next generation. This continuous till reaching the deadline. Finally, we evaluate the performance based on the GEBV of individuals in the final population.

\newpage
\subsection{Simulation Results}
To approximate the action-value function, we first generated learning data including state-action pairs using simulation and then trained a random forest algorithm for each generation separately to estimate the objective value. The size of training observations that were generated in the simulation vary between 1,500 to 6,000 for each generation, and there were a total 28 predictors including the action ($b$), maximum current GEBV ($g^{\max}$), remaining budget ($B$) and 25 values from the potential matrix, $C$. For training the random forest models, we did a search grid over three parameters including the number of selected features, minimum leaf node size, and maximum number of splits. The set of parameters with the least out of bag error were selected. The out of bag mean square errors for the first generation until the fourth generation are 2.39, 2.41, 2.31, 2.25, respectively.

We compared the optimal resource allocation strategy with the even allocation strategy (i.e., allocating resources equally across all generations). Three hundred independent simulations were conducted for each strategy using MATLAB (R2021-a).

\begin{figure}[H] \centering 
\includegraphics[width=15 cm]{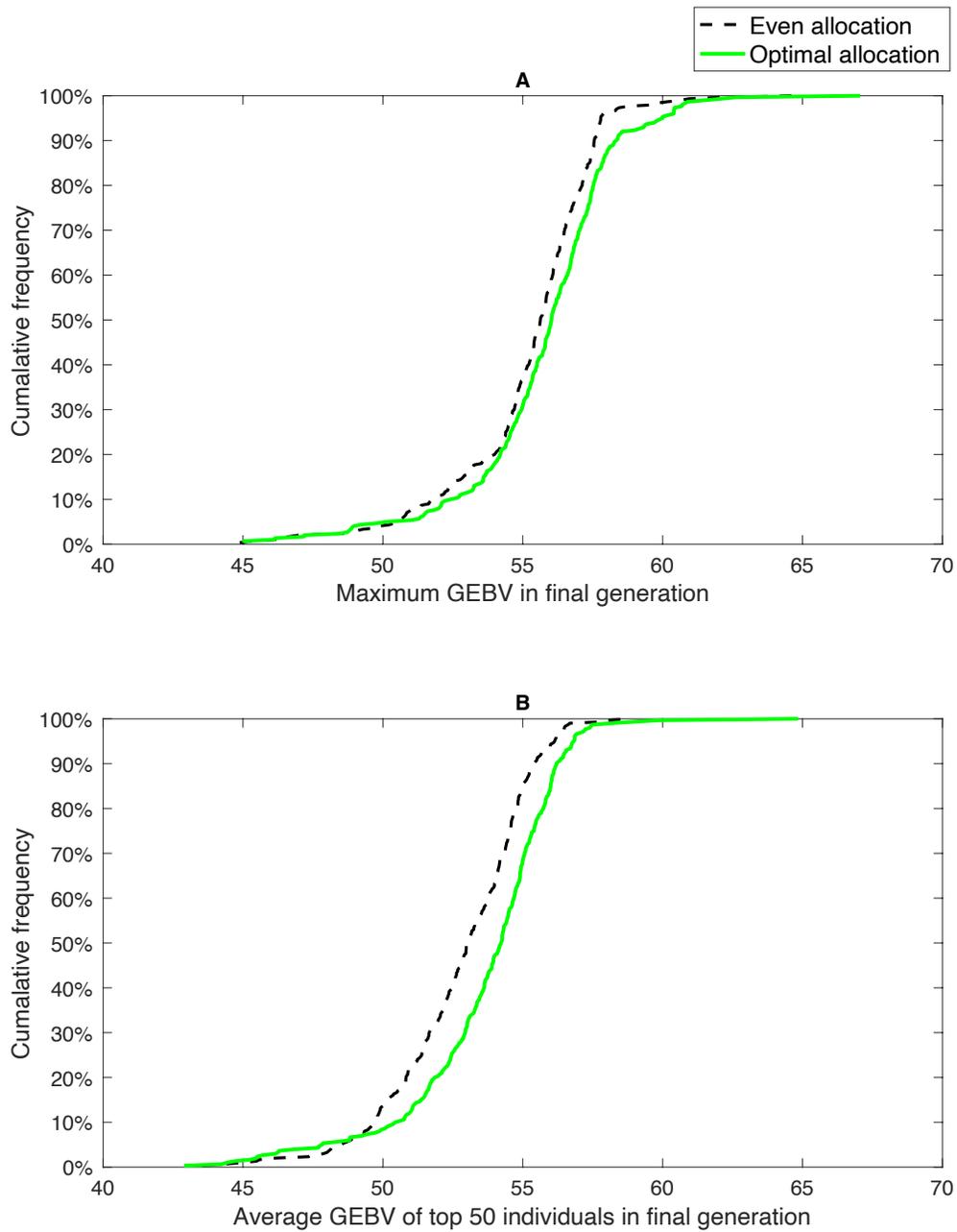}
\caption{Cumulative distribution functions of population maximum in the final generation (A) and average performance among top 50 individuals in the final generation (B) for two strategies of resource allocations among 300 independent simulations. The black dashed curve represents the even allocation strategy and the green curve represents the optimal allocation strategy.}
\label{CDF-RL}
\end{figure}

Figure \ref{CDF-RL} (A) demonstrates the cumulative distribution functions (CDFs) of the population maximum in the final generation. The performance becomes better as the CDF moves towards the right direction. Take for example, point (60, 92) means $92\%$ of the simulations achieved maximum GEBV less than or equal to 60. As demonstrated in this figure, the optimal allocation strategy outperforms even allocation strategy in almost all percentiles. Although, this improvement is not by a high margin, but it is considerable given that the improvement is across almost all percentiles for 5 generations of breeding. More improvements can be achieved for longer-term breeding. Moreover, if we compare the average performance of top 50 individuals instead of top 1, we can see a wider gap between the two curves as shown in Figure \ref{CDF-RL} (B). 

\begin{figure}[H] \centering 
\includegraphics[width=12.5cm, angle =90 ]{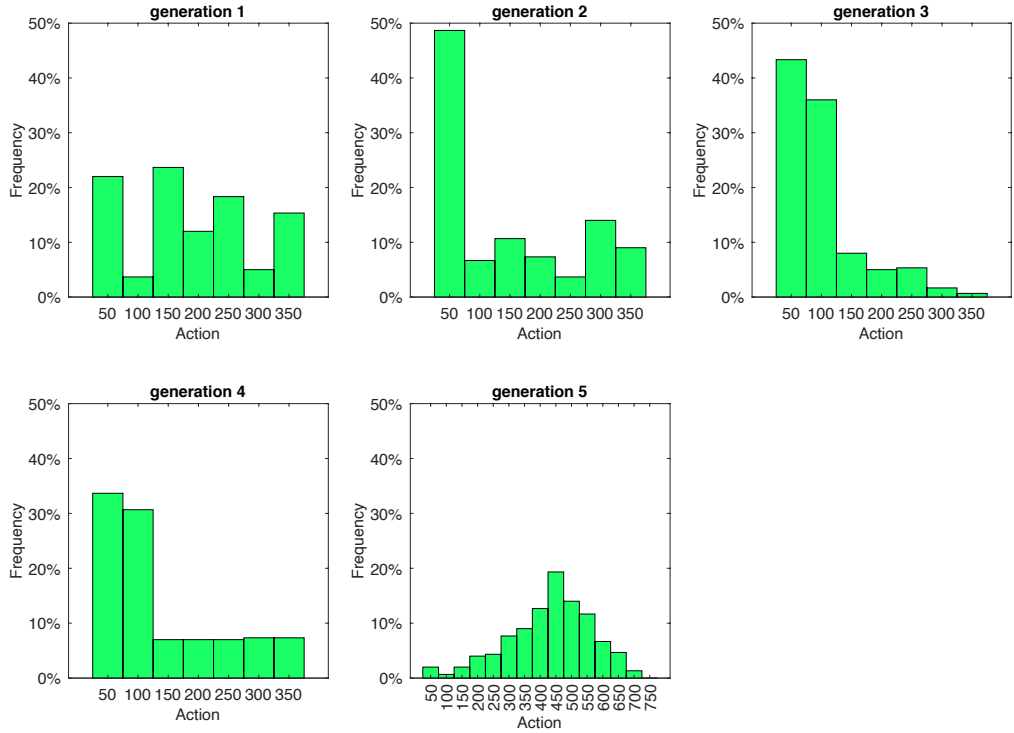}
\caption{Histograms of resource allocation across 5 generations for the optimal resource allocation strategy. The amount of resources that can be spent in all generations till one before final is chosen from a predefined set of actions. Here, we have seven different possibilities (50, 100, 150, 200, 250, 300, 350) for generations one till four and the remaining budget will be spent in the final generation. Note that for the even allocation strategy the action is deterministic which is $b=200$ for all generations.}
\label{hist-RL}
\end{figure}

So far, we have observed the improvements of our proposed optimal allocation strategy with respect to the even allocation strategy. Thus, the question arises: what is the behavior of the optimal allocation strategy and why that behavior results in improvements? To understand this better, we examined the histograms of resource allocation among 5 generations for the optimal strategy. As illustrated in Figure \ref{hist-RL}, three different behaviors are observed. In the first generation, the optimal strategy tends to be around the average, in the middle generations, there is more tendency towards spending less resources, and in the final generation almost half of the total budget is spent. 

In this case study, we demonstrated the effectiveness of our proposed method against evenly allocating resources across breeding cycles. Our optimal strategy suggests investing in the first generation and then spending moderate amount of resources in the middle generations and finally investing more in the final generation to exploit the best performance that can be achieved.

\section{Conclusion}
This study provides a framework to find the optimal resources that should be allocated throughout different generations in a breeding program by integrating the recently proposed look-ahead selection algorithm for genomic selection and reinforcement learning techniques.  Look-ahead selection is capable of estimating the consequences of selection and mating decisions under uncertain recombination events. Reinforcement learning is able to balance the trade-off between cost and time but its performance is sensitive to the definitions and dimensions of the state and action spaces. Therefore, look-ahead selection is integrated into the reinforcement learning framework to optimize resources in addition to the selection and mating steps and new solution techniques are proposed to battle the curse of dimensionality.

We considered MDPs with very large and continuous state spaces, and we used random forest to construct an approximate function to store the value functions used by the algorithms. We implemented a greedy policy improvement to learn optimal policy in a backward manner. Numerical results suggested the improvement of the proposed optimal allocation strategy versus even allocation strategy. 

The RL framework presented in this work has three major contributions. The first contribution is the definition of the state space. It is analytically and computationally challenging to simplify the state space definition for a large scale stochastic environment. To avoid the explosion of state space, we propose an integer linear program that captures the genomic information of the population by considering the trade-offs between time and resources. The second one is integrating the look-ahead selection and reinforcement learning. Given the optimal allocation strategy, look-ahead selection further improves the genetic gain by optimizing the selection and mating steps.
The third contribution is the learning process which is performed in a backward manner. We benefit from the structure of the genomic selection problem and assume we know the best policy in the target generation (spending all remaining budget). Then, we approximate the value function from the last generation to the first one and use it to improve the policy in a greedy way.

Future research is needed to address the limitations of this study. First, the current paper considers a sparsely discrete action space with predefined values. Future research should consider a more complete action space and investigate algorithms to optimize policy in such space. Second, deep neural networks can be used for function approximation if we generate more learning data by making the simulation more efficient. Finally, the case study presented here is for a single data set from a single crop organism. Future research considering more species is necessary to demonstrate the generalization of our proposed method.

\bibliographystyle{unsrtnat}
\bibliography{RLGS}  
\end{document}